\documentclass[utf8]{FrontiersinHarvard} 

\usepackage{url,hyperref,lineno,microtype,subcaption}
\usepackage[onehalfspacing]{setspace}
\usepackage[export]{adjustbox}
\usepackage{longtable}
\usepackage[T1]{fontenc}
\usepackage[utf8]{inputenc} 
\usepackage{newunicodechar}
\newunicodechar{́}{\'}
\usepackage{aas_macros}

\def\keyFont{\fontsize{8}{11}\helveticabold }
\def\firstAuthorLast{Liberato {et~al.}} 
\def\Authors{Giovanna Liberato\,$^{1,*}$, Denise R. Gonçalves\,$^{1}$, Arianna Cortesi\,$^{1,2}$, Luis Lomelí-Núñez\,$^{1}$, Alessandro Ederoclite\,$^{3,4}$,  Stavros Akras\,$^{5}$, Luis A. Gutiérrez-Soto\,$^{6,7}$, Vasiliki Fragkou\,$^{1}$, Marco Grossi\,$^{1}$, Eduardo Telles\,$^{8}$, Alvaro Alvarez-Candal\,$^{9}$, Fran Jiménez-Esteban\,$^{10}$, A.~J.~Cenarro\,$^{3}$, D.~Crist\'obal-Hornillos\,$^{3}$, C.~Hern\'andez-Monteagudo\,$^{11,12}$, C.~L\'opez-Sanjuan\,$^{3}$, A.~Mar\'{\i}n-Franch\,$^{3}$, M.~Moles\,$^{3}$, J.~Varela\,$^{3}$, H.~V\'azquez Rami\'o\,$^{3}$, R.~A.~Dupke\,$^{8,13}$, L.~Sodr\'e Jr.\,$^{14}$, R.~E.~Angulo\,$^{15,16}$ and the J-PLUS Collaboration}
\def\Address{$^{1}$Observatório do Valongo, Universidade Federal do Rio de Janeiro (OV-UFRJ) 
Ladeira Pedro Antonio 43, 20080-090 Rio de Janeiro, Brazil \\
$^{2}$Instituto de física, Universidade Federal do Rio de Janeiro (IF-UFRJ), 21941-972, Rio de Janeiro, Brazil \\
$^{3}$Centro de Estudios de Física del Cosmos de Aragón (CEFCA), Teruel, Spain \\
$^{4}$Unidad Asociada CEFCA-IAA, CEFCA, Unidad Asociada al CSIC por el IAA y el IFCA, Plaza San Juan 1, 44001 Teruel, Spain\label{UA}\\
$^{5}$Institute for Astronomy, Astrophysics, Space Applications and Remote Sensing, National Observatory of Athens (IAASARS-NOA), Athens, Greece \\
$^{6}$Instituto de Astrofísica de La Plata (CCT La Plata - CONICET - UNLP), B1900FWA, La Plata, Argentina \\
$^{7}$Instituto de F\'{i}sica, Universidade Federal do Rio Grande do Sul, Porto Alegre, RS 90040-060, Brazil\\
$^{8}$Observatório Nacional (ON), MCTI, Rua Gal. José Cristino 77, Rio de Janeiro 20921-400, RJ, Brazil \\
$^{9}$Instituto de Astrofísica de Andalucía, CSIC,  E18080 Granada, Spain \\
$^{10}$Centro de Astrobiologı́a (CAB), CSIC-INTA, Camino Bajo del Castillo s/n, E-28692, Villanueva de la Cañada, Madrid, Spain \\
$^{11}$Instituto de Astrof\'{\i}sica de Canarias, La Laguna, 38205, Tenerife, Spain\label{IAC} \\
$^{12}$Departamento de Astrof\'{\i}sica, Universidad de La Laguna, 38206, Tenerife, Spain\label{ULL} \\
$^{13}$University of Michigan, Department of Astronomy, 1085 South University Ave., Ann Arbor, MI 48109, USA\label{MU}\\
$^{14}$Instituto de Astronomia, Geof\'{\i}sica e Ci\^encias Atmosf\'ericas, Universidade de S\~ao Paulo, 05508-090 S\~ao Paulo, Brazil\label{USP}\\
$^{15}$Donostia International Physics Centre (DIPC), Paseo Manuel de Lardizabal 4, 20018 Donostia-San Sebastián, Spain\label{DIPC}\\
$^{16}$IKERBASQUE, Basque Foundation for Science, 48013, Bilbao, Spain\label{ikerbasque}
}
\def\corrAuthor{Giovanna Liberato}

\def\corrEmail{giovanna19@ov.ufrj.br}

\newcommand{\hii}{H~{\sc ii}}

\newcommand{\oi}{[O~{\sc i}]}

\newcommand{\oiii}{[O~{\sc iii}]}

\usepackage{hyperref}

\usepackage{color}

\begin{document}
\onecolumn
\firstpage{1}

\title[J-PLUS: The planetary nebula population of M~33]{J-PLUS: The planetary nebula population of M~33} 

\author[\firstAuthorLast ]{\Authors} 
\address{} 
\correspondance{} 

\extraAuth{}

\maketitle

\begin{abstract}
In this pilot study, we investigate the PN population in M~33, a nearby spiral galaxy ($\simeq 840$~kpc), using  data from the DR3 of the Javalambre-Photometric Local Universe Survey (J-PLUS), a 12-band photometric dataset extensively used to identify H$\alpha$ line emitters. 
From the 143 known PNe of M~33, the photometry of only 13 are present in the J-PLUS catalog, as available on the J-PLUS portal.
With the aim of recovering a larger fraction of the M~33 PN population, the software SExtractor is adopted to extract the sources in the J-PLUS images and obtain the photometric data for the PNe known in the literature, performing PSF photometry when possible. With this procedure the photometry of 98 PNe was obtained using H$\alpha$ image as detection image, including the 13 already present in the J-PLUS catalog.
Using diagnostic color-color diagrams (DCCDs) based on criteria developed for Milky Way halo PNe, we identified 16 sources with PN-like colors. Cross-match with existing catalogs revealed that most of these candidates are \hii~regions, though one source remains unidentified. Additionally, analyzing their full width at half maximum, most of them would not be PN candidates. This highlights the method's ability to select emission-line objects but also underscores
the challenge of distinguishing PNe from contaminants using photometry alone.
The J-PLUS colors of 98 known PNe were analyzed, together with literature information on their radial velocities, resulting in the identification of one possible halo PN.  
This is the first paper which aims at detecting extragalactic PNe in multi-band surveys such as J-PLUS, the Southern Local Universe Survey (S-PLUS) and the Javalambre-Physics of the Accelerating Universe Astrophysical Survey (J-PAS), paving the way for similar studies in these surveys for other nearby galaxies, which lack catalogs of known PNe.

\section{}


\tiny
 \keyFont{ \section{Keywords:} M~33, Planetary Nebulae, Multiband Surveys, Local Group, Photometry, Filters} 
\end{abstract}

\section{Introduction}

Low- and intermediate-mass stars, during their asymptotic giant branch evolutionary phases, undergo thermal pulses that eject significant amounts of their main sequence mass. The ejected circumstellar material is then ionized by the ultra-violet (UV) radiation from the hot stellar core, which results in the planetary nebulae (PNe) we observe. They are key objects because they can reveal information about: the chemistry of the interstellar medium from which the stellar progenitor was formed; the nucleosynthesis of stars in the above mentioned mass range; as well as the old stellar population kinematics. 
Furthermore, PNe can be found even in the outskirts of galaxies \citep{corradi15, Xiang17, galera-rosillo18}, and can be easily identified by their bright and characteristic emission lines, such as the Balmers series lines, \oi \,to \oiii, He I and He II, among others of abundant species of low- and intermediate ionization. Therefore, PNe are key tracers of the chemical and dynamical evolution of galaxies, as representative of the intermediate- and old-stellar population.

The M~33 galaxy (NGC~598) is a spiral galaxy in the Local Group of galaxies (LG), with the inclination ranging from 53° to 58° across the disk \citep{kam+17}. Along with M 31 and the Milky Way (MW), it is one of the three largest spiral galaxies in the LG, at approximately 840 kpc \citep{freedman}. At this distance,  the theoretical size of a typical PN ($\simeq$ 0.1~pc) is around 1'', therefore PNe can be considered unresolved, point-like sources, in M~33.

The PN population in M~33 has been studied over the past few decades \citep{magrini00, ciardullo04, galera-rosillo18}. In these works candidates were selected through narrow-band imaging targeting strong emission lines combined with broadband continuum images (e.g., B, V, g, r). These works  selected PN candidates based on their \oiii\ and H$\alpha$ excess, spatial compactness, lack of IR counterparts, and appropriate \oiii/H$\alpha$ flux ratios. \cite{magrini00} identified unresolved emission-line sources by subtracting scaled continuum images to remove the stellar background of M~33, and selecting only those with negligible continuum; they obtained the optical spectra of 39 PN candidates and confirmed 26 of them as very likely genuine PNe. \cite{ciardullo04} used a blinking technique to compare \oiii\ and H$\alpha$ images with V-band frames, selecting unresolved objects with strong emission lines and very low continuum; they observed spectroscopically in the optical and confirmed 140 of 152 candidates, including the ones identified by \cite{magrini00}. \cite{galera-rosillo18} performed a deep wide-field search using both visual inspection of color-composite images (\oiii/g and H$\alpha$/r) and photometric analysis via color-color diagrams, confirming 3 from 8 candidates through optical spectroscopy. To date, the joint previous efforts resulted in 143 spectroscopically confirmed PNe in M~33, within a galactocentric distance of 40~kpc. 

An increasing number of recent photometric surveys make use of broad-, intermediate-, and narrow-band filters to address many scientific questions without the need for spectroscopy, which offers a more efficient way to observe and characterize millions of astronomical sources across large areas of the sky. One of these surveys is the Javalambre Photometric Local Universe Survey \citep[J-PLUS,][]{cenarro19}, which has been extensively used to search for and study H$\alpha$ emitters \citep{vilella-rojo15, logrono19, gutierrez-soto19, lumbreras-calle22, Rahna25, gutierrez-soto25}. \cite{gutierrez-soto19} use J-PLUS and S-PLUS colors of {\sc cloudy} models representative of the halo PN population in the Galaxy and propose new photometric diagnostic criteria to identify compact PNe in this component of the Milky Way.

The aim of the present work is to explore J-PLUS data to extract a catalogue of PNe in M~33, using the sample of PNe from the literature as reference, and analyze their properties using J-PLUS colors through diagnostic tools such as color–color diagrams (DCCD). 
By analyzing the extracted sources and PNe through the criteria developed by \cite{gutierrez-soto19} to search for halo PNe (hPNe) in the Galaxy, as well as the morphology (size) of those objects, we can also investigate if a fraction of these PNe may belong to M~33's halo and asses the possibility that some of the extracted sources might be PN candidates.
The paper is structure as follows: in Section \ref{sec:catalog} we present the methodology developed to extract the PNe catalogues in the 12 J-PLUS bands, using SExtractor and PSFEx. In Section \ref{sec:caracterization}
we characterize the PNe population using the J-PLUS filter system and kinematics from the literature. Finally, in Section \ref{sec:discussion} we interpret the results and we discuss them in the context of multi-band surveys possibility of extracting catalogue of PN candidates.


\section{J-PLUS imagery and photometric catalog}
\label{sec:catalog}

\begin{figure}[h!]
\begin{center}
\includegraphics[width=0.9\linewidth]{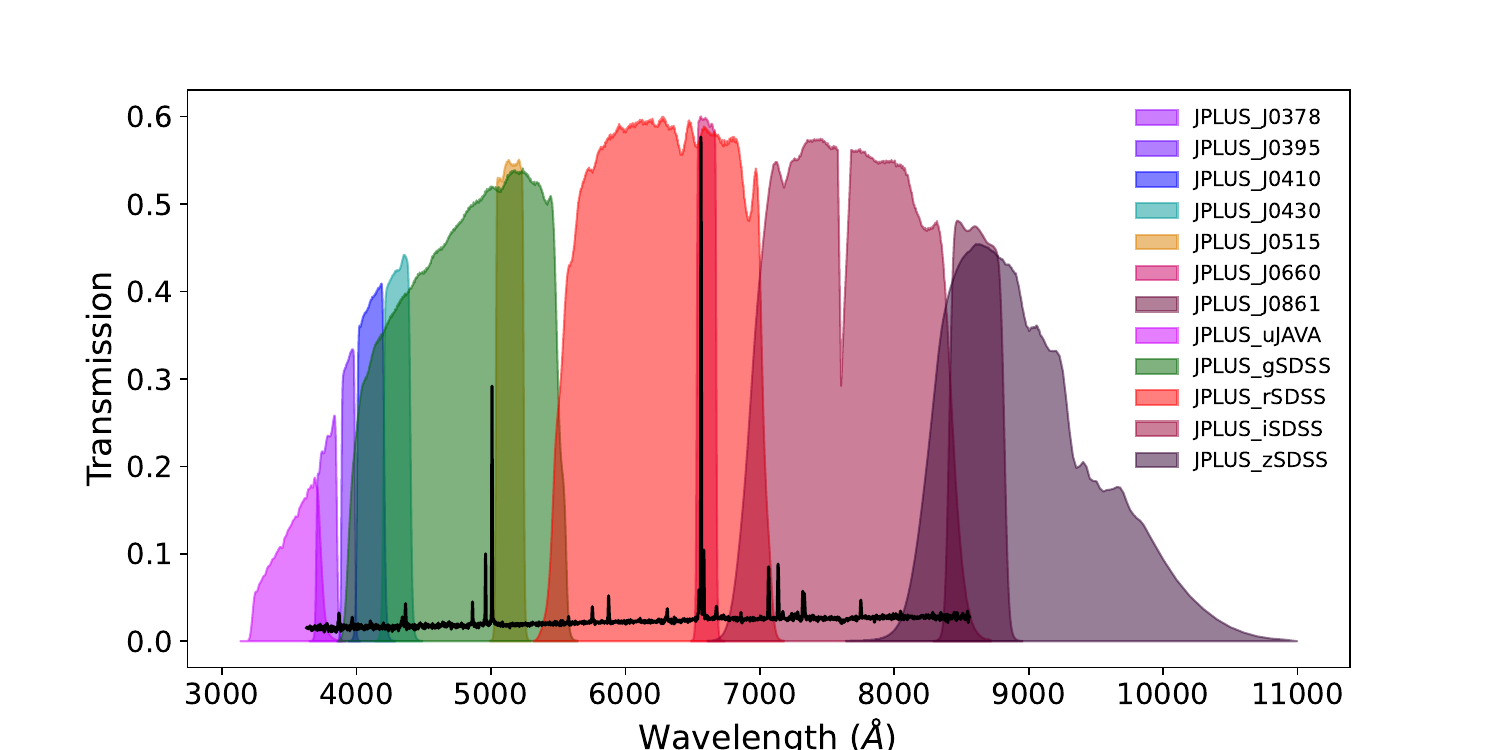}
\end{center}
\caption{J-PLUS transmission curves superposed with the spectrum of a planetary nebula selected by VPHAS+ and ALLWISE color criteria and spectroscopically confirmed 
\citep[][see their Fig.~4]{liberato+23}. It is possible to see the H$\alpha$ emission in the spectrum (black line) falling over the $J0660$ filter, as indicated in the label. The \oiii\ duplet is also visible, at the edge of the filter $J0515$.}\label{jplus-filters}
\end{figure}

J-PLUS is an ongoing photometric survey based on 12-bands (5 broad and 7 narrow), in the optical range, from 3500~\AA\ to 10,000~\AA. The survey observes the Northern Hemisphere from the Observatorio Astrofísico
de Javalambre (OAJ; \citealt{oaj}) with the 83 cm Javalambre Auxiliary Survey Telescope (JAST80) and the $9200 \times 9200$ pixels camera T80Cam \citep{t80cam}. J-PLUS third data release (DR3) comprises 1642 fields which amounts to 3284 deg$^{2}$, with seeing ($\simeq 1"$) and the pixel scale of 0.56"/pixel \citep{lopez-sanjuan+24}.
The 12-filter system is sensitive both to the continuum and to important spectral features such as the emission lines present in the PN spectra, which gives us the opportunity to study PNe in a multi-band fashion. The J-PLUS filter system, as well as the optical spectrum of a PN, can be seen in Fig. \ref{jplus-filters}.

\begin{figure}[h!]
\begin{center}
\includegraphics[width=0.7\linewidth]{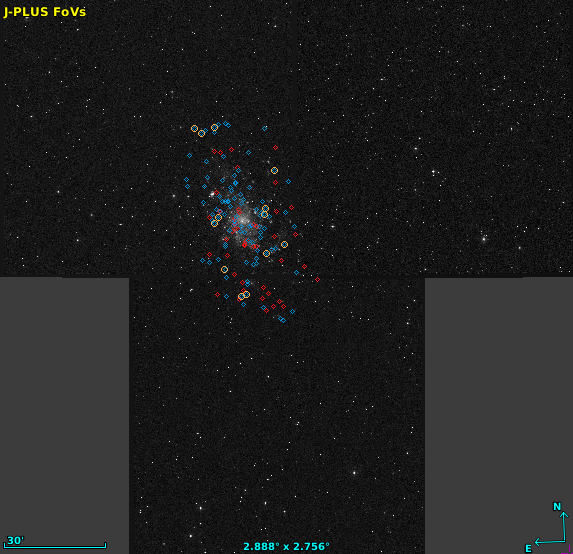}
\end{center}
\caption{The three H$\alpha$ J-PLUS fields of view selected to cover the M~33 galaxy. J-PLUS images 04327, 04326 and 04283, with central coordinates (RA,Dec) of 23.9935,31.1133, upper left), 22.3818,31.1133, upper right and 23.2496,29.7209, bottom, respectively. Each image is $\approx$ 2 deg$^{2}$. The markers represent the 143 PNe, being in blue the 98 PNe we recovered, in red the 45 missing and the 13 ones already catalogued on J-PLUS are marked in yellow. North is up, east to the left.}\label{jplus-fields}
\end{figure}

From the cross-match of the total sample of 143 PNe in M~33 compiled from the literature \citep{magrini00, ciardullo04, galera-rosillo18} and the J-PLUS DR3 catalogue\footnote{https://archive.cefca.es/catalogues/jplus-dr3}, 13 PNe were identified. In order to reveal the rest of the M~33 PNe, possibly hidden in the J-PLUS images, we extract the sources and measure their photometry in the 12-band images, as described in Section \ref{sec:sextractor}. 
We thus work with 36 images comprised in the three J-PLUS fields of view (FoVs) needed to cover M~33 (images 04283, 04326 and 04327 from J-PLUS database). The three J-PLUS FoVs can be seen in Fig. \ref{jplus-fields}.

\subsection{Point-sources detection and photometry in the J-PLUS images}
\label{sec:sextractor}
SExtractor \citep{bertin96} and  PSFEx \citep{bertin11} were used to recover as many as possible of the known PNe and their photometry in M~33. The former, performs the extraction and the photometry of individual sources in astronomical images via: the estimation of the sky background, thresholding, separating close sources that have been extracted as single sources, filtering of the detections, photometry and star/galaxy separation. In order to extract the individual sources, SExtractor considers a number of parameters that specify the image properties,
the source detection threshold, the photometric aperture, and the zero point magnitude. The physical quantities to be computed (including magnitudes, fluxes, coordinates and associated errors) also need to be defined, per run. We used SExtractor in its dual mode, on which the source detection is performed in one image and the extraction is done in all the other images. PSFEx --an optimized, fully automated, two-dimensional model-fitting code-- is implemented to build PSF models based on the  point-like sources detected by SExtractor. We selected the most appropriated sources for the PSFEx models, those that are point sources, and with a satisfactory model  created, we rerun SExtractor to extract a sources-catalogue with PSF photometry measured.
 
\begin{figure}[h!]
\begin{center}
\includegraphics[width=0.75\linewidth]{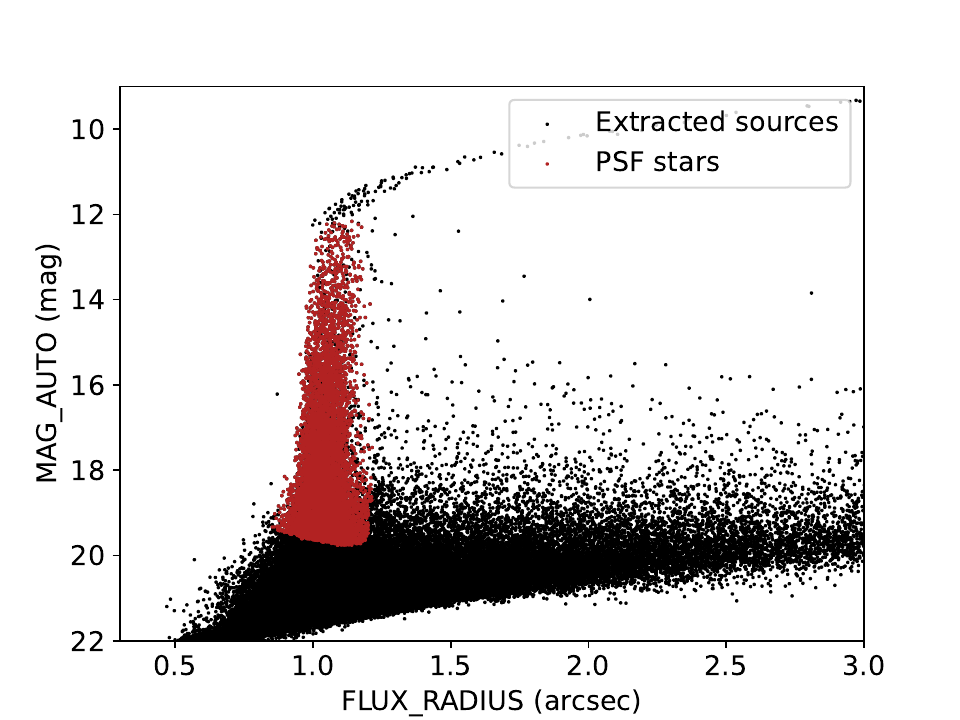}
\end{center}
\caption{MAG\_AUTO vs FLUX\_RADIUS space used to select point-sources for the PSF model. In red are shown the selected ones, leaving out the nearly horizontal line in the upper part of the plot that can comprise saturated sources. The population of fainter magnitudes are extended or spurious sources, like cosmic rays.}
\label{psf-stars}
\end{figure}

SExtractor was executed for all the 12-band images across the three fields using the dual mode with the H$\alpha$ as the detection image,
since it was the one for which the recovered number of PNe was maximized. Table~\ref{tab:parameters} shows the values of the most sensitive parameters among the SExtractor-run configurations, which are the same for all bands, with the exception of MAG\_ZEROPOINT (the magnitude's zero point) whose values, per filter, were extracted from the information provided by the J-PLUS survey, as it can be seen on table \ref{tab:magzeropoint}. The remaining parameters were set to their default values. To run PSFEx, we selected the stars based on the parameter space MAG\_AUTO $\times$ FLUX\_RADIUS, making sure to leave out saturated or extended sources and cosmic rays. The selected stars to build the PSF model for one of the 36 images are shown in red in Fig. \ref{psf-stars}, as an example, and the PSFEx sensitive parameters are shown in Table \ref{tab:parameters} as well. In the second run of SExtractor to obtain the PSF photometry, we use a 3-pixel ($\approx$ 1.7 
") diameter aperture, since the average full width at half maximum (FWHM) of the extracted PNe is $\approx$ 2.7 pixels ($\approx$ 1.5").

\subsection{PNe recovery and photometry}

Following the procedure described in Section \ref{sec:sextractor}, more than 100,000 sources were extracted and among them 98 known PNe. Fig.~\ref{jplus-fields} shows the 143 known PNe in M~33 marked as red squares, the 13 PNe present in the J-PLUS catalogue are marked by a yellow circle, and the 98 PNe recovered in this work can be seen in blue squares that lie on top of the red ones. We note that the 13 M~33 PNe that are cataloged in the J-PLUS DR3 were also successfully recovered by the procedure of the present work. We can also see from this figure that the sample of PNe with J-PLUS photometry was expanded, both in the central regions and in the outskirts of the galaxy. This demonstrates that J-PLUS has great potential for tracing PNe, although its automatic source extraction is not fully optimized to detect them efficiently.

The sample of 98 known PNe of M~33 extracted by SExtractor is presented with the complementary materials of this paper. The sources' coordinates, aperture ($\approx$ 3.7 arcsec) and PSF magnitudes and magnitude errors for the 12 filters as well as the signal to noise ratios are listed in the complementary table. For a fraction of the sources, neither the PSF nor the aperture magnitude was well measured in all filters.
To analyze the recovered PNe sample, we use the PSF magnitudes when available or the 3" aperture magnitude instead.

\begin{figure}[h!]
\begin{center}
\includegraphics[width=0.7\linewidth]{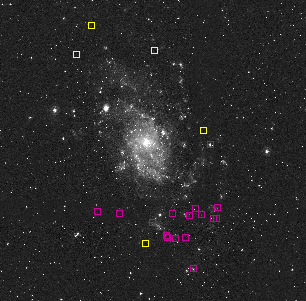}
\end{center}
\caption{H$\alpha$ M~33 mosaic image made with the three J-PLUS fields. The markers represent extracted sources that pass through the hPNe criteria for the MW halo \cite{gutierrez-soto19} that will be discussed in Section 3.1, being the yellow ones confirmed PNe \citep{magrini00, ciardullo04, galera-rosillo18} and the magenta possible PN candidates. North is up, east to the left. The image has a size of $\approx$ 1 degree.}\label{m33sources}
\end{figure}

\section{Characterization of the M~33 PN population}
\label{sec:caracterization}

With our sample of 98 known PNe in M~33, we 
conduct a more detailed analysis of these objects, exploring the novel J-PLUS filters.
We, thus, complement the J-PLUS photometric data with kinematic data, available in \cite{ciardullo04}, to investigate the possible presence of halo PNe, and the relation between kinematics and colors.

\subsection{J-PLUS diagnostic color-color diagrams}
\label{sec:color-color}

\begin{figure}[h!]
\begin{center}
\includegraphics[width=0.7\linewidth]{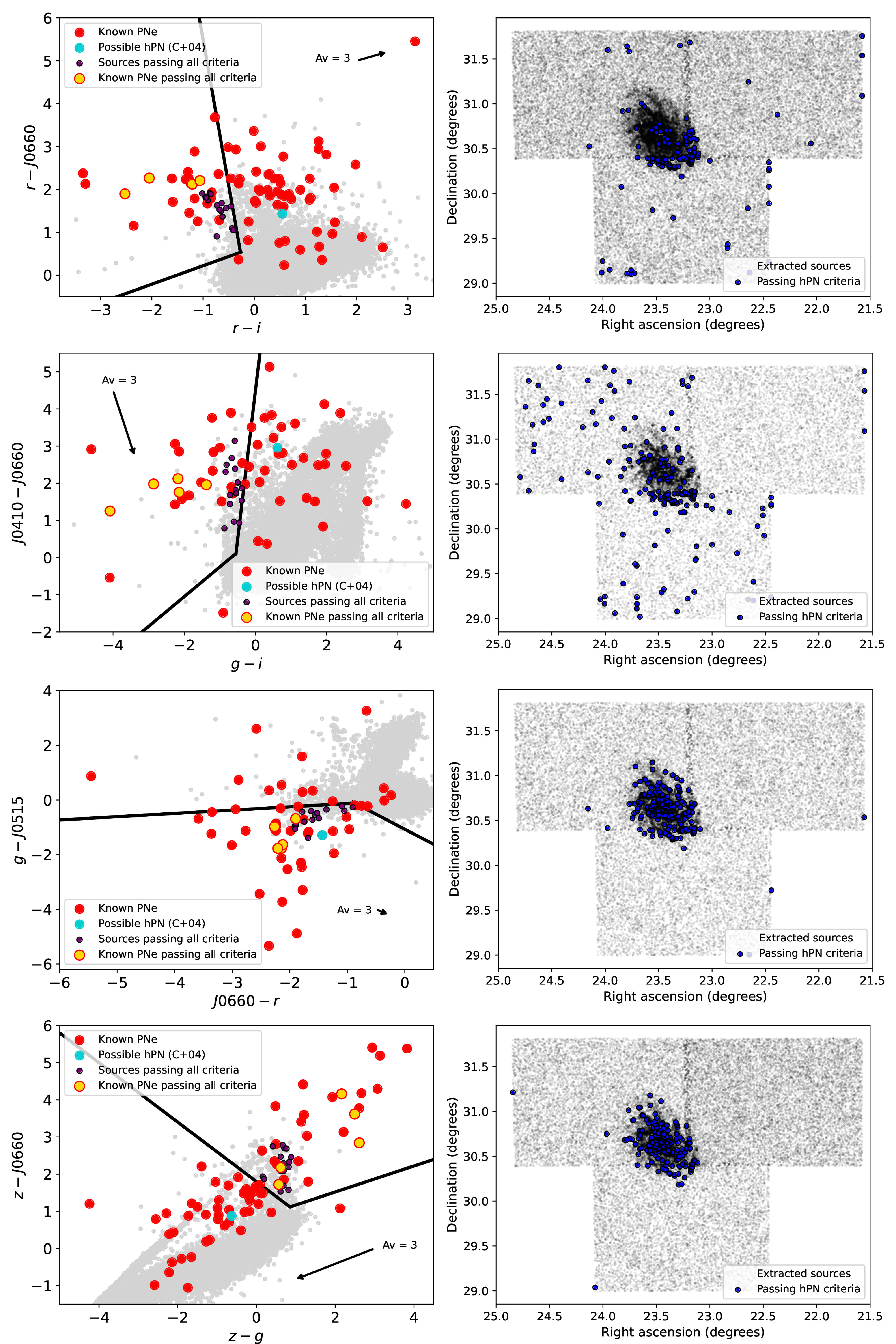}
\end{center}
\caption{Left: DCCDs showing the extracted sources from the field images in gray and in dark purple the sources that passes all the criteria from \cite{gutierrez-soto19}, represented as black lines. The extracted PNe are shown in red, except for 5 of them in yellow representing the ones that pass all the criteria at the same time, and the one in light blue representing one PN that according to \cite{ciardullo04} might be a halo PNe. Right: Distribution of the selected objects in each DCCDs.}\label{DCCDs}
\end{figure}

In order to characterize the PN population of the nearby galaxy M~33 with the J-PLUS photometry, we make use of diagnostic color-color diagrams (DCCD).
This approach allows us to explore their photometric properties, compare PNe with other types of emission-line objects, such as symbiotic stars, H~II regions, T-Tauri stars, and Herbig-Haro and young stellar objects, and investigate possible trends or outliers that may provide insights about their nature. In this work, we investigate the PN population in 4 different DCCDs, r-i vs r-J0660, g-i vs J0410-J0660, J0660-r vs g-J0515, and z-g vs z-J0660, identified by the collaboration as useful to separate PNe within meaningful intervals, as proposed by \cite{gutierrez-soto19}. One of the DCCDs proposed by these authors is left out the analysis due to the fact it is extinction dependent. These DCCDs can be seen in Figure~\ref{DCCDs}.
For each DCCD, the number of PNe is limited to the availability of the particular magnitudes under the analyses, a situation that also applies to sources other than PNe.
The total number of objects per diagram is shown in Table~\ref{tab:CCD}. In Figure~\ref{DCCDs}, the black lines represent the color criteria developed by \cite{gutierrez-soto19} to search for Galactic hPNe. The criteria can be found on Table \ref{tab:Hcriteria}.
An upper limit of 0.2 magnitude was applied to the errors of the 12 magnitudes of the extracted sources. We note that no cut in error was applied to the known PNe.

The left panel of Fig. \ref{DCCDs} presents the DCCDs showing the known PN population of M~33 (in red), the one among them that according to \cite{ciardullo04} possibly belongs to M~33's halo (in light blue), the 16 SExtractor extracted sources that pass the four MW hPNe criteria by \cite{gutierrez-soto19} here discussed (in purple), the 5 known PNe that also pass these criteria (yellow) and the other extracted sources in the J-PLUS FoVs (in gray). On the right panel, we show the distribution of the objects selected by the color criteria of each diagram.
It is possible to notice that the objects criteria based on the i band return  more sources in the southern part of the galaxy, located on the top of the bottom image. This is probably due to different observations conditions. These DCCDs enlighten the fact that several known PNe have similar characteristics as the Milky Way hPNe. This is somewhat expected, since the chemical abundances of the M~33 PNe should better compare with those of the Galactic halo, than the higher-metallicity MW disk. 
The PN population of M~33 is characterized by 12+log(O/H)=8.42 \citep{magrini+09}, and the bulk of this population is composed by PNe of the disk. In fact, as shown by \cite{barker07} and \cite{McConnachie06} the stellar halo of the galaxy contributes for less than a few percent to M~33 total luminosity, which is in agreement with \cite{ciardullo04} finding of only two {\it possible} halo PNe in this galaxy, of which only one was recovered in our photometry. 
It is worthy emphasizing that the MW hPNe in \cite{gutierrez-soto19} have abundances of 7.61$<$12+log(O/H)$<$8.19, a range that is closer to the chemical state of the disk PNe in M~33 than to that of the MW disk \citep[8.69,][]{perinotto+04}. The former and latter arguments, justify our findings of about as many known PNe obeying the halo criteria as tracing other components of galaxy. Therefore, what these DCCDs are actually tracing are PNe chemically poorer than the Galactic disk PNe.

As for the 16 sources marked by dark purple circles on the DCCDs, their positions in these diagrams suggest they are strong emission-line objects. We further investigate their nature in Sec. \ref{sec:discussion}, where a cross-match with literature catalogs reveals that while the selection criteria are effective at identifying emission-line sources, they do not cleanly separate PNe from other object classes such as compact \hii~regions and globular clusters.
However, known PNe (red and yellow circles) are relatively more present than other sources (gray dots) within the black lines that define the MW halo PNe in \citet{gutierrez-soto19}, rendering these regions of interest for identifying possible candidates for PN or emission line objects.

\begin{figure}[h!]
\begin{center}
\includegraphics[width=0.45\linewidth]{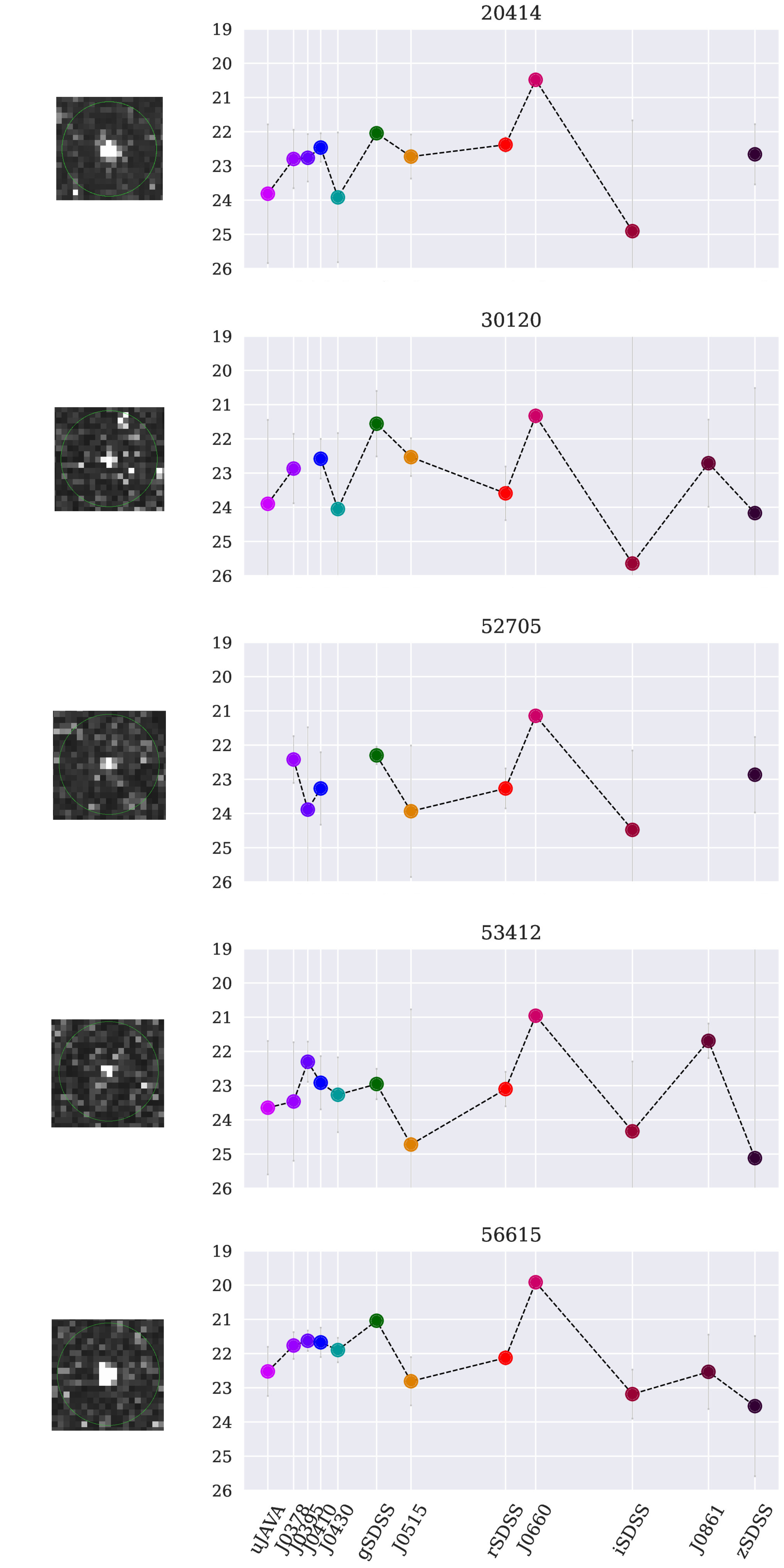}
\end{center}
\caption{J-PLUS SEDs of the 5 literature PNe passing the criteria from \cite{gutierrez-soto19}}.  
\label{seds-pn}
\end{figure}

 The 5 known PNe passing the hPNe color criteria, as well as the likely halo PN, are discussed in terms of their radial velocities, therefore their possibility of actually belonging to the M~33's halo, in Section \ref{sec:kinematics}. The J-PLUS SEDs of the 5  hPNe and the 16 extracted sources passing the criteria can be found in Figs. \ref{seds-pn} and \ref{16seds}, respectively.

\subsection{Kinematics}
\label{sec:kinematics}

\begin{figure}[h!]
\begin{center}
\includegraphics[width=0.7\linewidth]{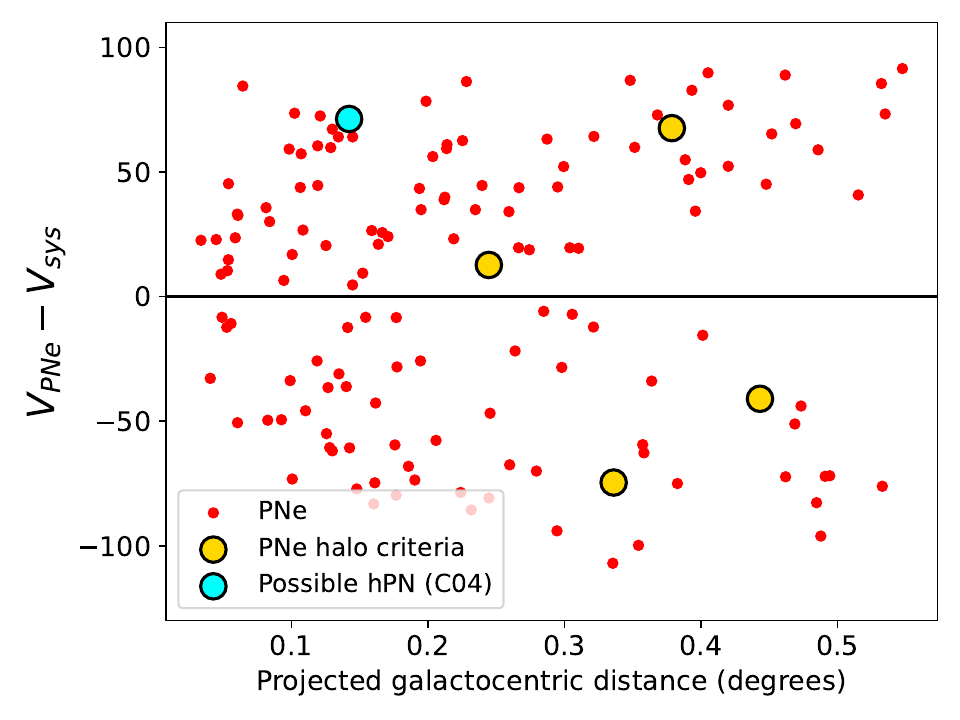}
\end{center}
\caption{Phase space diagram showing the difference between the PNe velocities and the systemic velocity vs the galactocentic distance of the PNe. The four yellow symbols represent the PNe passing all the MW halo criteria from \cite{gutierrez-soto19} and the one in cyan represents one PN that according to \cite{ciardullo04} might be in the M~33's halo.}
\label{velocities}
\end{figure}

Different galaxy components, i.e. disk, bulge and halo, 
exhibit different kinematic profiles. For example, the motion of stars, as PNe, in the disk is supported by rotation, with low dispersion velocities \citep{cortesi+11}, while the bulge and halo are pressure supported, and therefore dominated by random motions. Recovering the kinematics of the PNe is a powerful tool to determine which components they belong to. In Fig.~\ref{velocities} we present the one-dimensional phase-space diagram, i.e. the difference between the PN velocities from \cite{ciardullo04} and the systemic velocity of M~33 versus the projected PN galactocentric distances.
In a first approximation, stars belonging to the disk will occupy the upper and lower region of this plot, while halo objects, whose velocity is close to the systemic velocity  \citep[given that the dispersion velocity is 20~km/s,][]{ciardullo04}, would lie along the filled black line and near the galaxy center. The yellow filled circles represent four of the five PNe that passed the selection DCCD criteria for which we have velocity estimations, and the PN suggested as possible halo PN by \cite{ciardullo04} in cyan. 
From Fig.~\ref{velocities}, only one of the halo-like PNe would be actually associated to the M~33's halo according to this simple model. 
It is clear that the selection of possible hPNe using only DCCDs differs from the one using only kinematics. These findings show the importance of combining kinematics and stellar population properties to properly derive a galaxy formation history. A more realistic kinematic model should be developed in a future work.

\section{Discussion and conclusions}
\label{sec:discussion}

In this work, we have investigated the PN population of M~33 using J-PLUS photometry. By extracting sources from J-PLUS images and measuring their photometry with SExtractor and PSFEx, we successfully recovered photometric data for 98 previously known PNe, compared to only 13 objects originally listed in the J-PLUS catalog. In this way, we recovered more than two-thirds of the known PN population of M~33. This method fails to recover 31.5\% of the known PN sample due to a variety of reasons; for example, objects located in the crowded central regions of M~33, or on top of spiral arms or \hii~regions, which can make it difficult to extract the PNe. This demonstrates that J-PLUS data, although not specifically optimized for PNe detection, can be used effectively to recover and characterize extragalactic PN populations. We have also explored the use of combined images (r,i,H$\alpha$ and g,r,i) to optimize our ability of objects finding. While the total number of recovered known PNe was 65, therefore lower than when using the H$\alpha$ image as detection image, we also identified additional 9 PNe. In further studies we will explore the possibility of combining the two methods to improve our results. 

\begin{figure}[h!]
\begin{center}
\includegraphics[width=0.85\linewidth]{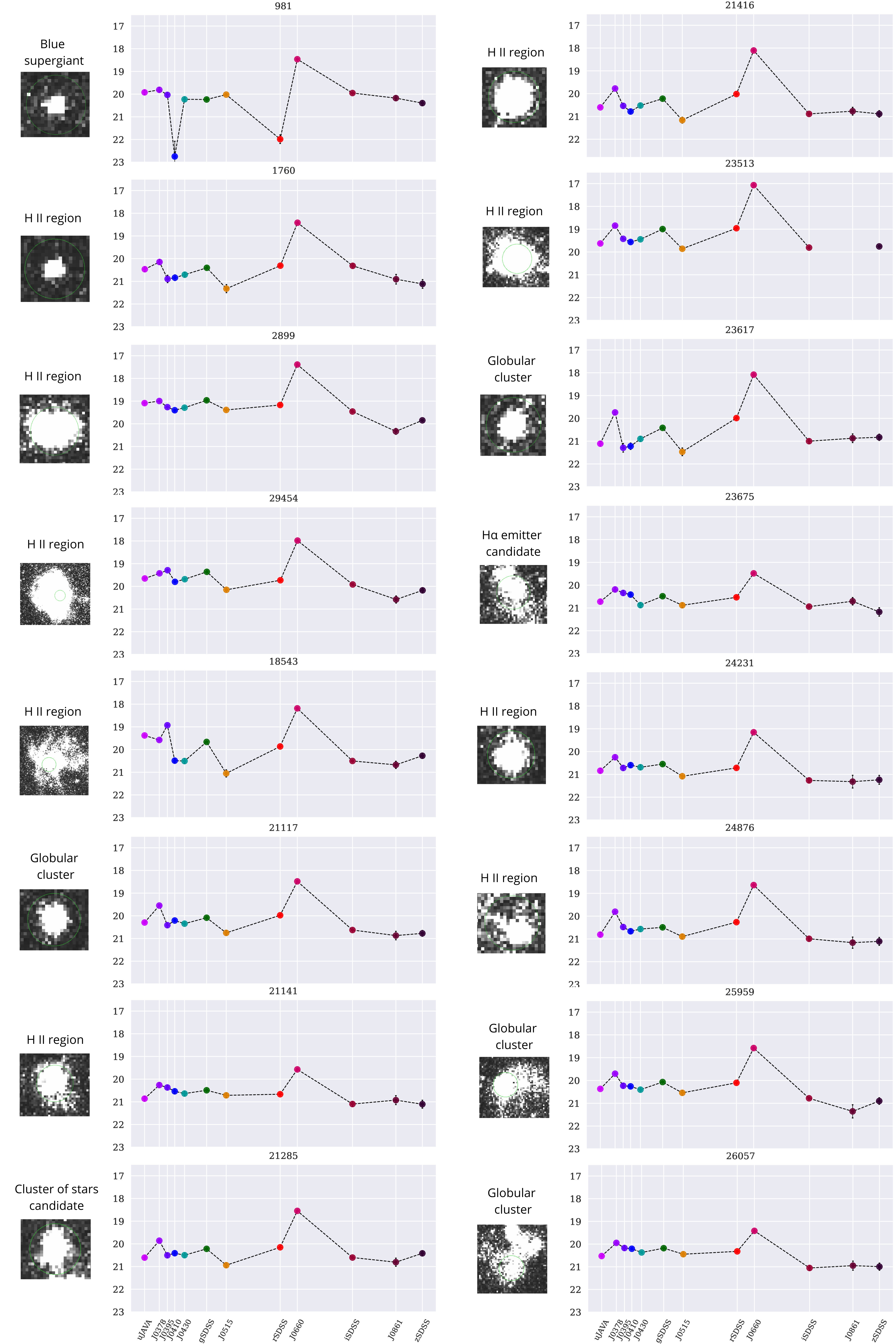}
\end{center}
\caption{SEDs of the 16 extracted sources passing all criteria from \cite{gutierrez-soto19}. The green circle in the object images has a 0.08 radius.}.  
\label{16seds}
\end{figure}

The analysis of the individual DCCDs shows that many of the known M~33 PNe have some photometric properties similar to the halo PNe of the Milky Way, which is consistent with the intermediate metallicity of M~33. Our analysis of the DCCDs (Fig. \ref{DCCDs}), using color criteria from \cite{gutierrez-soto19}, demonstrated that many known M~33 PNe
occupy similar color space to Milky Way halo PNe, consistent with M~33's intermediate metallicity. However, when these criteria were
applied to our full source catalog, the method proved to have low purity. We identified 16 sources that satisfied all color cuts, but analyzing their FWHM measured by SExtractor, they are extended sources. If we apply a limit of 3 pixel ($\approx$ 1.7") to the FWHM, according to the average FWHM of the PNe ($\approx$ 2.7 pixel, or 1.5"), only one of those objects would remain a candidate, the first one in Fig. \ref{16seds}. This shows the importance of combining color criteria with some sort of size criteria as well. 

As we can see in Fig. \ref{16seds}, literature search revealed that the vast majority are previously identified objects: 9 \hii~regions, 4 globular clusters, one blue supergiant and one star cluster candidate. Only one candidate lacks prior identification in the literature. This one candidate has a high value of FWHM measured by SExtractor, however, performing a visual inspection, the object seems to be overlapping a more extended source. This high contamination rate demonstrates that while the J-PLUS DCCDs are effective at selecting emission-line objects, additional criteria -- likely requiring the \oiii~ filter planned for J-PAS or spectroscopic follow-up -- are necessary to reliably distinguish bona fide PNe from compact \hii~regions and other contaminants. However, when we compare the SEDs of the PNe with the ones of the objects identified as \hii~regions, we notice that the emission in the g band relatively to the two adjacent J-PLUS bands (J-0430 and J0515) is stronger for PNe.  
This can be attributed to the presence of the intense \oiii~ emission lines within the g band wavelength range. The PNe show higher \oiii~ lines because of their very hot central stars; in contrast, \hii~regions, which are ionized by cooler massive stars, show weaker \oiii~ emission and are instead dominated by recombination lines such as H$\alpha$, resulting in a lower relative g band fluxes.

Regrading the  \hii~regions, and more in general the 16 sources, selected through the color criteria from \cite{gutierrez-soto19}, we note that they are mostly detected in the southern outskirts of the galaxy (return to Fig. \ref{jplus-fields} to see their location), corresponding to the bottom tile. We tentatively attribute this result to the different depths of the filters in the three tiles, arising from inhomogeneous observing conditions, as discussed in Appendix A of \cite{lomeli25}.

When combined with the available kinematic information, our analysis indicates that only one of the PNe passing the halo color criteria could be associated with M~33’s stellar halo. This result is in agreement with previous works showing that the stellar halo contributes very little to the galaxy’s luminosity. Therefore, the bulk of the PN population traced by J-PLUS belongs to the disk of M~33.

This work also provides a valuable complement to the recent, comprehensive census of resolved stellar populations in M~33. The
Panchromatic Hubble Andromeda Treasury: Triangulum Extended Region (PHATTER) survey has provided an unprecedentedly deep, high-resolution view of the stars that constitute M~33, mapping its star formation history and structure \citep{williams21}. While PHATTER provides the definitive context of the parent stellar populations, our J-PLUS study isolates a key endpoint of low- and intermediate-mass stellar evolution: the planetary nebula phase. By linking our catalogue of PNe to the underlying properties of the stellar disk and halo as mapped by PHATTER, we can directly probe the connection between a star's birth environment and its ultimate fate.
This synergy offers a powerful path to investigating how factors like age and metallicity, derived from PHATTER, influence the
characteristics of PN populations.

These results show the strengths of using multiband photometric surveys to identify PNe candidates, opening a promising field of research in the J-PAS era, which will complement the H$\alpha$
filter already present in J-PLUS with the \oiii\ one, among others, as well as being $\simeq 2$ mag deeper than J-PLUS. With these new J-PLUS available data for most of the PNe on M~33, future works can make use of them to, for example, constrain physical properties of PNe through multiband SED fitting or to derive selection criteria for PNe candidates, enabling more complete samples in nearby galaxies.

\renewcommand{\arraystretch}{1.5}
\begin{table}[]
\centering
\caption{Sensitive parameters of SExtractor and PSFEx used in our runs.}
\vspace{0.2cm}
\label{tab:parameters}
\begin{tabular}{cc}
\hline
\multicolumn{2}{c}{SExtractor parameters configuration} \\ \hline
\footnotesize DETECT\_MINAREA & \footnotesize 2 \\
\footnotesize DETECT\_THRESH & \footnotesize 1.5 \\
\footnotesize ANALYSIS\_THRESH & \footnotesize 1.5 \\
\footnotesize BACK\_SIZE & \footnotesize 64 \\
\footnotesize PIXEL\_SCALE & \footnotesize 0.5556 \\ \hline
\multicolumn{2}{c}{PSFEx parameters configuration} \\ \hline\footnotesize
PSF\_SIZE & \footnotesize 25,25 \\
\footnotesize PSFVAR\_DEGREES & \footnotesize 3
\end{tabular}
\end{table}
\renewcommand{\arraystretch}{1}

\renewcommand{\arraystretch}{1.5}
\begin{table}[]
\centering
\caption{Number of PNe and extracted sources for each DCCD in Fig.~\ref{DCCDs}.}
\vspace{0.2cm}
\label{tab:CCD}
\begin{tabular}{ccc}
\hline
DCCD & PNe & \multicolumn{1}{l}{Extracted sources} \\ \hline
\textit{r-i vs r-J0660} & 77 & 112468 \\
g-i vs J0410-J0660 & 54 & 88458 \\
J0660-r vs g-J0515 & 51 & 100217 \\
z-g vs z-J0660 & 71 & 108731
\end{tabular}
\end{table}

\begin{table}[ht]
\centering
\caption{Color criteria of each DCCD from \cite{gutierrez-soto19}.}
\vspace{0.3cm}
\label{tab:Hcriteria}
\begin{tabular}{|c|c|}
\hline
Diagrams & Criteria \\ \hline
r-i vs r-J0660 & \begin{tabular}[c]{@{}c@{}}(r - J0660) $\geq$ 0.43 $\times$ (r - i) + 0.65 \&\\ (r - J0660) $\leq$ -6.80 $\times$ (r - i) - 1.30\end{tabular} \\ \hline
g-i vs J0410-J0660 & \begin{tabular}[c]{@{}c@{}}(J0410 - J0660) $\geq$ 8.0 $\times$ (g - i) + 4.50 \&\\ (J0410 - J0660) $\geq$ 0.8 $\times$ (g - i) + 0.55\end{tabular} \\ \hline
J0660-r vs g-J0515 & \begin{tabular}[c]{@{}c@{}}(g - J0515) $\leq$ 0.12 $\times$ (J0660 - r) - 0.01 \&\\ (g - J0515) $\leq$ -1.10 $\times$ (J0660 - r) - 1.07\end{tabular} \\ \hline
z-g vs z-J0660 & \begin{tabular}[c]{@{}c@{}}(z - J0660) $\geq$ 0.35 $\times$ (z - g) + 0.82 \&\\ (z - J0660) $\geq$ -0.80 $\times$ (z - g) + 1.80\end{tabular} \\ \hline
\end{tabular}
\end{table}

\begin{table}[]
\centering
\caption{Zero point magnitudes for each field of view and each filter used on SExtractor to measure source magnitudes.}
\vspace{0.3cm}
\label{tab:magzeropoint}
\begin{tabular}{|c|c|c|c|}
\hline
\multicolumn{1}{|l|}{Filter} & \multicolumn{1}{l|}{Image 04283} & \multicolumn{1}{l|}{Image 04326} & \multicolumn{1}{l|}{Image 04327} \\ \hline
u & 21.0948 & 21.272 & 20.9469 \\ \hline
J0378 & 20.4583 & 20.6171 & 20.3387 \\ \hline
J0395 & 20.3074 & 20.46 & 20.2709 \\ \hline
J0410 & 21.2898 & 21.425 & 21.2433 \\ \hline
J0430 & 21.352 & 21.4777 & 21.2723 \\ \hline
g & 23.5768 & 23.6826 & 23.5334 \\ \hline
J0515 & 21.5509 & 21.6355 & 21.4968 \\ \hline
r & 23.6352 & 23.7124 & 23.571 \\ \hline
J0660 & 21.087 & 21.1649 & 21.0148 \\ \hline
i & 23.3607 & 23.4016 & 23.2387 \\ \hline
J0861 & 21.6437 & 21.6835 & 21.5432 \\ \hline
z & 22.8523 & 22.8073 & 22.7776 \\ \hline
\end{tabular}
\end{table}

\renewcommand{\arraystretch}{1}

\section*{Conflict of Interest Statement}

The authors declare that the research was conducted in the absence of any commercial or financial relationships that could be construed as a potential conflict of interest.

\section*{Author Contributions}

GL: Conduction of the research, obtaining the data, analyses and writing.
DRG and AC: Project organization, supervision,
advisement, writing, analyses and review.
LLN: Helped with SExtractor and PSFEx softwares.
AE, SA, LAGS, VF and MG offered valuable comments on photometry, M~33, planetary nebulae limiting magnitudes and J-PLUS photometry.
ET, AAC and FJE as members of J-PLUS offered significant contributions to improve the text.


\section*{Funding}
Funding for the J-PLUS Project has been provided by the Governments of Spain and Arag\'on through the Fondo de Inversiones de Teruel; the Aragonese Government through the Research Groups E96, E103, E16\_17R, E16\_20R, and E16\_23R; the Spanish Ministry of Science and Innovation (MCIN/AEI/10.13039/501100011033 y FEDER, Una manera de hacer Europa) with grants PID2021-124918NB-C41, PID2021-124918NB-C42, PID2021-124918NA-C43, and PID2021-124918NB-C44; the Spanish Ministry of Science, Innovation and Universities (MCIU/AEI/FEDER, UE) with grants PGC2018-097585-B-C21 and PGC2018-097585-B-C22; the Spanish Ministry of Economy and Competitiveness (MINECO) under AYA2015-66211-C2-1-P, AYA2015-66211-C2-2, AYA2012-30789, and ICTS-2009-14; and European FEDER funding (FCDD10-4E-867, FCDD13-4E-2685). The Brazilian agencies FINEP, FAPESP, and the National Observatory of Brazil have also contributed to this Project.

\section*{Acknowledgments}
GL’s research is supported by an M.Sc. student’s grant (88887.948597/2024-00) from CAPES – the Brazilian Federal Agency for Support and Evaluation of Graduate Education within the Education Ministry. DRG acknowledges the grants of the Brazilian agencies FAPERJ (E-26/200.527/2023) and CNPq (403011/2022-1; 315307/2023-4). 
AC aknowledges the FAPERJ grants E26/202.607/2022 and 210.371/2022(270993) and CNPq.
AE acknowledges the financial support from the Spanish Ministry of Science and Innovation and the European Union - NextGenerationEU through the Recovery and Resilience Facility project ICTS-MRR-2021-03-CEFCA. VF thanks FAPERJ for granting the postdoctoral research fellowships E-26/200.181/2025 and 200.181/2025(304980). MG acknowledges support from FAPERJ grant E-26/211.370/2021. AAC acknowledges financial support from the Severo Ochoa grant CEX2021-001131-S and the project PID2023-153123NB-I00 funded by MCIN/AEI/10.13039/501100011033 and the FSE+.
This research has partially funded by MICIU/AEI/10.13039/501100011033/ through grant PID2023-146210NB-I00.

Based on observations made with the JAST80 telescope and T80Cam camera for the J-PLUS project at the Observatorio Astrof\'{\i}sico de Javalambre (OAJ), in Teruel, owned, managed, and operated by the Centro de Estudios de F\'{\i}sica del  Cosmos de Arag\'on (CEFCA). We acknowledge the OAJ Data Processing and Archiving Department (DPAD; \citealt{upad}) for reducing and calibrating the OAJ data used in this work.



\bibliographystyle{Frontiers-Harvard} 
\bibliography{test}

@ARTICLE{magrini00,
       author = {{Magrini}, L. and {Corradi}, R.~L.~M. and {Mampaso}, A. and {Perinotto}, M.},
        title = "{A search for planetary nebulae in M33}",
      journal = {\aap},
         year = 2000,
        month = mar,
       volume = {355},
        pages = {713-719},
       adsurl = {https://ui.adsabs.harvard.edu/abs/2000A&A...355..713M}
}

@ARTICLE{ciardullo04,
       author = {{Ciardullo}, Robin and {Durrell}, Patrick R. and {Laychak}, Mary Beth and {Herrmann}, Kimberly A. and {Moody}, Kenneth and {Jacoby}, George H. and {Feldmeier}, John J.},
        title = "{The Planetary Nebula System of M33}",
      journal = {\apj},
         year = 2004,
        month = oct,
       volume = {614},
       number = {1},
        pages = {167-185},
          doi = {10.1086/423414},
archivePrefix = {arXiv},
       eprint = {astro-ph/0406462},
 primaryClass = {astro-ph},
       adsurl = {https://ui.adsabs.harvard.edu/abs/2004ApJ...614..167C}
}

@ARTICLE{galera-rosillo18,
       author = {{Galera-Rosillo}, R. and {Corradi}, R.~L.~M. and {Mampaso}, A.},
        title = "{A deep narrowband survey for planetary nebulae at the outskirts of M 33}",
      journal = {\aap},
         year = 2018,
        month = apr,
       volume = {612},
          eid = {A35},
        pages = {A35},
          doi = {10.1051/0004-6361/201731383},
archivePrefix = {arXiv},
       eprint = {1712.07595},
 primaryClass = {astro-ph.GA},
       adsurl = {https://ui.adsabs.harvard.edu/abs/2018A&A...612A..35G}
}

@INPROCEEDINGS{Xiang17,
       author = {{Xiang}, Maosheng and {Liu}, Xiaowei and {Zhang}, Meng and {Yuan}, Haibo and {Huo}, Zhiying},
        title = "{The LAMOST spectroscopic survey of planetary nebulae in M31 and M33}",
    booktitle = {Planetary Nebulae: Multi-Wavelength Probes of Stellar and Galactic Evolution},
         year = 2017,
       editor = {{Liu}, X. and {Stanghellini}, L. and {Karakas}, A.},
       series = {IAU Symposium},
       volume = {323},
        month = oct,
        pages = {388-389},
          doi = {10.1017/S1743921317002356},
       adsurl = {https://ui.adsabs.harvard.edu/abs/2017IAUS..323..388X}
}

@ARTICLE{corradi15,
       author = {{Corradi}, R.~L.~M. and {Kwitter}, K.~B. and {Balick}, B. and {Henry}, R.~B.~C. and {Hensley}, K.},
        title = "{The Chemistry of Planetary Nebulae in the Outer Regions of M31}",
      journal = {\apj},
         year = 2015,
        month = jul,
       volume = {807},
       number = {2},
          eid = {181},
        pages = {181},
          doi = {10.1088/0004-637X/807/2/181},
archivePrefix = {arXiv},
       eprint = {1504.03869},
 primaryClass = {astro-ph.GA},
       adsurl = {https://ui.adsabs.harvard.edu/abs/2015ApJ...807..181C}
}

@ARTICLE{cenarro19,
       author = {{Cenarro}, A.~J. and {Moles}, M. and {Crist{\'o}bal-Hornillos}, D. and {Mar{\'\i}n-Franch}, A. and {Ederoclite}, A. and {Varela}, J. and {L{\'o}pez-Sanjuan}, C. and {Hern{\'a}ndez-Monteagudo}, C. and {Angulo}, R.~E. and {V{\'a}zquez Rami{\'o}}, H. and {Viironen}, K. and {Bonoli}, S. and {Orsi}, A.~A. and {Hurier}, G. and {San Roman}, I. and {Greisel}, N. and {Vilella-Rojo}, G. and {D{\'\i}az-Garc{\'\i}a}, L.~A. and {Logro{\~n}o-Garc{\'\i}a}, R. and {Gurung-L{\'o}pez}, S. and {Spinoso}, D. and {Izquierdo-Villalba}, D. and {Aguerri}, J.~A.~L. and {Allende Prieto}, C. and {Bonatto}, C. and {Carvano}, J.~M. and {Chies-Santos}, A.~L. and {Daflon}, S. and {Dupke}, R.~A. and {Falc{\'o}n-Barroso}, J. and {Gon{\c{c}}alves}, D.~R. and {Jim{\'e}nez-Teja}, Y. and {Molino}, A. and {Placco}, V.~M. and {Solano}, E. and {Whitten}, D.~D. and {Abril}, J. and {Ant{\'o}n}, J.~L. and {Bello}, R. and {Bielsa de Toledo}, S. and {Castillo-Ram{\'\i}rez}, J. and {Chueca}, S. and {Civera}, T. and {D{\'\i}az-Mart{\'\i}n}, M.~C. and {Dom{\'\i}nguez-Mart{\'\i}nez}, M. and {Garzar{\'a}n-Calderaro}, J. and {Hern{\'a}ndez-Fuertes}, J. and {Iglesias-Marzoa}, R. and {I{\~n}iguez}, C. and {Jim{\'e}nez Ruiz}, J.~M. and {Kruuse}, K. and {Lamadrid}, J.~L. and {Lasso-Cabrera}, N. and {L{\'o}pez-Alegre}, G. and {L{\'o}pez-Sainz}, A. and {Ma{\'\i}cas}, N. and {Moreno-Signes}, A. and {Muniesa}, D.~J. and {Rodr{\'\i}guez-Llano}, S. and {Rueda-Teruel}, F. and {Rueda-Teruel}, S. and {Soriano-Lagu{\'\i}a}, I. and {Tilve}, V. and {Valdivielso}, L. and {Yanes-D{\'\i}az}, A. and {Alcaniz}, J.~S. and {Mendes de Oliveira}, C. and {Sodr{\'e}}, L. and {Coelho}, P. and {Lopes de Oliveira}, R. and {Tamm}, A. and {Xavier}, H.~S. and {Abramo}, L.~R. and {Akras}, S. and {Alfaro}, E.~J. and {Alvarez-Candal}, A. and {Ascaso}, B. and {Beasley}, M.~A. and {Beers}, T.~C. and {Borges Fernandes}, M. and {Bruzual}, G.~R. and {Buzzo}, M.~L. and {Carrasco}, J.~M. and {Cepa}, J. and {Cortesi}, A. and {Costa-Duarte}, M.~V. and {De Pr{\'a}}, M. and {Favole}, G. and {Galarza}, A. and {Galbany}, L. and {Garcia}, K. and {Gonz{\'a}lez Delgado}, R.~M. and {Gonz{\'a}lez-Serrano}, J.~I. and {Guti{\'e}rrez-Soto}, L.~A. and {Hernandez-Jimenez}, J.~A. and {Kanaan}, A. and {Kuncarayakti}, H. and {Landim}, R.~C.~G. and {Laur}, J. and {Licandro}, J. and {Lima Neto}, G.~B. and {Lyman}, J.~D. and {Ma{\'\i}z Apell{\'a}niz}, J. and {Miralda-Escud{\'e}}, J. and {Morate}, D. and {Nogueira-Cavalcante}, J.~P. and {Novais}, P.~M. and {Oncins}, M. and {Oteo}, I. and {Overzier}, R.~A. and {Pereira}, C.~B. and {Rebassa-Mansergas}, A. and {Reis}, R.~R.~R. and {Roig}, F. and {Sako}, M. and {Salvador-Rusi{\~n}ol}, N. and {Sampedro}, L. and {S{\'a}nchez-Bl{\'a}zquez}, P. and {Santos}, W.~A. and {Schmidtobreick}, L. and {Siffert}, B.~B. and {Telles}, E. and {Vilchez}, J.~M.},
        title = "{J-PLUS: The Javalambre Photometric Local Universe Survey}",
      journal = {\aap},
         year = 2019,
        month = feb,
       volume = {622},
          eid = {A176},
        pages = {A176},
          doi = {10.1051/0004-6361/201833036},
archivePrefix = {arXiv},
       eprint = {1804.02667},
 primaryClass = {astro-ph.GA},
       adsurl = {https://ui.adsabs.harvard.edu/abs/2019A&A...622A.176C}
}

@ARTICLE{gutierrez-soto19,
       author = {{Guti{\'e}rrez-Soto}, L.~A. and {Gon{\c{c}}alves}, D.~R. and {Akras}, S. and {Cortesi}, A. and {L{\'o}pez-Sanjuan}, C. and {Guerrero}, M.~A. and {Daflon}, S. and {Borges Fernandes}, M. and {Mendes de Oliveira}, C. and {Ederoclite}, A. and {Sodr{\'e}}, L. and {Pereira}, C.~B. and {Kanaan}, A. and {Werle}, A. and {V{\'a}zquez Rami{\'o}}, H. and {Alcaniz}, J.~S. and {Angulo}, R.~E. and {Cenarro}, A.~J. and {Crist{\'o}bal-Hornillos}, D. and {Dupke}, R.~A. and {Hern{\'a}ndez-Monteagudo}, C. and {Mar{\'\i}n-Franch}, A. and {Moles}, M. and {Varela}, J. and {Ribeiro}, T. and {Schoenell}, W. and {Alvarez-Candal}, A. and {Galbany}, L. and {Jim{\'e}nez-Esteban}, F.~M. and {Logro{\~n}o-Garc{\'\i}a}, R. and {Sobral}, D.},
        title = "{J-PLUS: Tools to identify compact planetary nebulae in the Javalambre and southern photometric local Universe surveys}",
      journal = {\aap},
         year = 2020,
        month = jan,
       volume = {633},
          eid = {A123},
        pages = {A123},
          doi = {10.1051/0004-6361/201935700},
archivePrefix = {arXiv},
       eprint = {1912.10145},
 primaryClass = {astro-ph.GA},
       adsurl = {https://ui.adsabs.harvard.edu/abs/2020A&A...633A.123G}
}

@ARTICLE{Rahna25,
       author = {{Rahna}, P.~T. and {Akhlaghi}, M. and {L{\'o}pez-Sanjuan}, C. and {Logro{\~n}o-Garc{\'\i}a}, R. and {Muniesa}, D.~J. and {Dom{\'\i}nguez-S{\'a}nchez}, H. and {Fern{\'a}ndez-Ontiveros}, J.~A. and {Sobral}, D. and {Lumbreras-Calle}, A. and {Chies-Santos}, A.~L. and {Rodr{\'\i}guez-Mart{\'\i}n}, J.~E. and {Eskandarlou}, S. and {Ederoclite}, A. and {Alvarez-Candal}, A. and {V{\'a}zquez Rami{\'o}}, H. and {Cenarro}, A.~J. and {Mar{\'\i}n-Franch}, A. and {Alcaniz}, J. and {Angulo}, R.~E. and {Crist{\'o}bal-Hornillos}, D. and {Dupke}, R.~A. and {Hern{\'a}ndez-Monteagudo}, C. and {Moles}, M. and {Sodr{\'e}}, L. and {Varela}, J.},
        title = "{J-PLUS: Spectroscopic validation of H{\ensuremath{\alpha}} emission line maps in spatially resolved galaxies}",
      journal = {\aap},
         year = 2025,
        month = mar,
       volume = {695},
          eid = {A200},
        pages = {A200},
          doi = {10.1051/0004-6361/202453633},
archivePrefix = {arXiv},
       eprint = {2502.05830},
 primaryClass = {astro-ph.GA},
       adsurl = {https://ui.adsabs.harvard.edu/abs/2025A&A...695A.200R}
}

@ARTICLE{lumbreras-calle22,
       author = {{Lumbreras-Calle}, A. and {L{\'o}pez-Sanjuan}, C. and {Sobral}, D. and {Fern{\'a}ndez-Ontiveros}, J.~A. and {V{\'\i}lchez}, J.~M. and {Hern{\'a}n-Caballero}, A. and {Akhlaghi}, M. and {D{\'\i}az-Garc{\'\i}a}, L.~A. and {Alcaniz}, J. and {Angulo}, R.~E. and {Cenarro}, A.~J. and {Crist{\'o}bal-Hornillos}, D. and {Dupke}, R.~A. and {Ederoclite}, A. and {Hern{\'a}ndez-Monteagudo}, C. and {Mar{\'\i}n-Franch}, A. and {Moles}, M. and {Sodr{\'e}}, L. and {V{\'a}zquez Rami{\'o}}, H. and {Varela}, J.},
        title = "{J-PLUS: Uncovering a large population of extreme [OIII] emitters in the local Universe}",
      journal = {\aap},
         year = 2022,
        month = dec,
       volume = {668},
          eid = {A60},
        pages = {A60},
          doi = {10.1051/0004-6361/202142898},
archivePrefix = {arXiv},
       eprint = {2112.06938},
 primaryClass = {astro-ph.GA},
       adsurl = {https://ui.adsabs.harvard.edu/abs/2022A&A...668A..60L}
}

@ARTICLE{logrono19,
       author = {{Logro{\~n}o-Garc{\'\i}a}, R. and {Vilella-Rojo}, G. and {L{\'o}pez-Sanjuan}, C. and {Varela}, J. and {Viironen}, K. and {Muniesa}, D.~J. and {Cenarro}, A.~J. and {Crist{\'o}bal-Hornillos}, D. and {Ederoclite}, A. and {Mar{\'\i}n-Franch}, A. and {Moles}, M. and {V{\'a}zquez Rami{\'o}}, H. and {Bonoli}, S. and {D{\'\i}az-Garc{\'\i}a}, L.~A. and {Orsi}, A. and {San Roman}, I. and {Akras}, S. and {Chies-Santos}, A.~L. and {Coelho}, P.~R.~T. and {Daflon}, S. and {Costa-Duarte}, M.~V. and {Dupke}, R. and {Galbany}, L. and {Gonz{\'a}lez Delgado}, R.~M. and {Hernandez-Jimenez}, J.~A. and {Lopes de Oliveira}, R. and {Mendes de Oliveira}, C. and {Oteo}, I. and {Gon{\c{c}}alves}, D.~R. and {S{\'a}nchez-Portal}, M. and {Schmidtobreick}, L. and {Sodr{\'e}}, L.},
        title = "{J-PLUS: Measuring H{\ensuremath{\alpha}} emission line fluxes in the nearby universe}",
      journal = {\aap},
         year = 2019,
        month = feb,
       volume = {622},
          eid = {A180},
        pages = {A180},
          doi = {10.1051/0004-6361/201732487},
archivePrefix = {arXiv},
       eprint = {1804.04039},
 primaryClass = {astro-ph.GA},
       adsurl = {https://ui.adsabs.harvard.edu/abs/2019A&A...622A.180L}
}

@ARTICLE{vilella-rojo15,
       author = {{Vilella-Rojo}, G. and {Viironen}, K. and {L{\'o}pez-Sanjuan}, C. and {Cenarro}, A.~J. and {Varela}, J. and {D{\'\i}az-Garc{\'\i}a}, L.~A. and {Crist{\'o}bal-Hornillos}, D. and {Ederoclite}, A. and {Mar{\'\i}n-Franch}, A. and {Moles}, M.},
        title = "{Extracting H{\ensuremath{\alpha}} flux from photometric data in the J-PLUS survey}",
      journal = {\aap},
         year = 2015,
        month = aug,
       volume = {580},
          eid = {A47},
        pages = {A47},
          doi = {10.1051/0004-6361/201526374},
archivePrefix = {arXiv},
       eprint = {1505.07115},
 primaryClass = {astro-ph.GA},
       adsurl = {https://ui.adsabs.harvard.edu/abs/2015A&A...580A..47V}
}

@ARTICLE{freedman,
       author = {{Freedman}, Wendy L. and {Wilson}, Christine D. and {Madore}, Barry F.},
        title = "{New Cepheid Distances to Nearby Galaxies Based on BVRI CCD Photometry. II. The Local Group Galaxy M33}",
      journal = {\apj},
         year = 1991,
        month = may,
       volume = {372},
        pages = {455},
          doi = {10.1086/169991},
       adsurl = {https://ui.adsabs.harvard.edu/abs/1991ApJ...372..455F}
}

@ARTICLE{bertin96,
       author = {{Bertin}, E. and {Arnouts}, S.},
        title = "{SExtractor: Software for source extraction.}",
      journal = {\aaps},
         year = 1996,
        month = jun,
       volume = {117},
        pages = {393-404},
          doi = {10.1051/aas:1996164},
       adsurl = {https://ui.adsabs.harvard.edu/abs/1996A&AS..117..393B}
}

@INPROCEEDINGS{bertin11,
       author = {{Bertin}, E.},
        title = "{Automated Morphometry with SExtractor and PSFEx}",
    booktitle = {Astronomical Data Analysis Software and Systems XX},
         year = 2011,
       editor = {{Evans}, I.~N. and {Accomazzi}, A. and {Mink}, D.~J. and {Rots}, A.~H.},
       series = {Astronomical Society of the Pacific Conference Series},
       volume = {442},
        month = jul,
        pages = {435},
       adsurl = {https://ui.adsabs.harvard.edu/abs/2011ASPC..442..435B}
}

@ARTICLE{lopez-sanjuan+24,
       author = {{L{\'o}pez-Sanjuan}, C. and {V{\'a}zquez Rami{\'o}}, H. and {Xiao}, K. and {Yuan}, H. and {Carrasco}, J.~M. and {Varela}, J. and {Crist{\'o}bal-Hornillos}, D. and {Tremblay}, P. -E. and {Ederoclite}, A. and {Mar{\'\i}n-Franch}, A. and {Cenarro}, A.~J. and {Coelho}, P.~R.~T. and {Daflon}, S. and {del Pino}, A. and {Dom{\'\i}nguez S{\'a}nchez}, H. and {Fern{\'a}ndez-Ontiveros}, J.~A. and {Hern{\'a}n-Caballero}, A. and {Jim{\'e}nez-Esteban}, F.~M. and {Alcaniz}, J. and {Angulo}, R.~E. and {Dupke}, R.~A. and {Hern{\'a}ndez-Monteagudo}, C. and {Moles}, M. and {Sodr{\'e}}, L.},
        title = "{J-PLUS: Toward a homogeneous photometric calibration using Gaia BP/RP low-resolution spectra}",
      journal = {\aap},
         year = 2024,
        month = mar,
       volume = {683},
          eid = {A29},
        pages = {A29},
          doi = {10.1051/0004-6361/202346012},
archivePrefix = {arXiv},
       eprint = {2301.12395},
 primaryClass = {astro-ph.SR},
       adsurl = {https://ui.adsabs.harvard.edu/abs/2024A&A...683A..29L}
}

@ARTICLE{liberato+23,
       author = {{Liberato}, Giovanna and {Goncalves}, D.~R. and {Gutierrez-Soto}, L.~A. and {Akras}, S.},
        title = "{Photometric candidate selection and spectroscopic confirmation of new PNe and SySts in the Galactic plane}",
      journal = {Bulletin of the Astronomical Society of Brazil},
         year = 2023,
        month = jan,
       volume = {34},
        pages = {100-103},
          doi = {10.48550/arXiv.2310.15335},
archivePrefix = {arXiv},
       eprint = {2310.15335},
 primaryClass = {astro-ph.GA},
       adsurl = {https://ui.adsabs.harvard.edu/abs/2023BASBr..34..100L}
}

@ARTICLE{kam+17,
       author = {{Kam}, S.~Z. and {Carignan}, C. and {Chemin}, L. and {Foster}, T. and {Elson}, E. and {Jarrett}, T.~H.},
        title = "{H I Kinematics and Mass Distribution of Messier 33}",
      journal = {\aj},
         year = 2017,
        month = aug,
       volume = {154},
       number = {2},
          eid = {41},
        pages = {41},
          doi = {10.3847/1538-3881/aa79f3},
       adsurl = {https://ui.adsabs.harvard.edu/abs/2017AJ....154...41K}
}

@ARTICLE{cortesi+11,
       author = {{Cortesi}, A. and {Merrifield}, M.~R. and {Arnaboldi}, M. and {Gerhard}, O. and {Martinez-Valpuesta}, I. and {Saha}, K. and {Coccato}, L. and {Bamford}, S. and {Napolitano}, N.~R. and {Das}, P. and {Douglas}, N.~G. and {Romanowsky}, A.~J. and {Kuijken}, K. and {Capaccioli}, M. and {Freeman}, K.~C.},
        title = "{Unravelling the origins of S0 galaxies using maximum likelihood analysis of planetary nebulae kinematics}",
      journal = {\mnras},
         year = 2011,
        month = jun,
       volume = {414},
       number = {1},
        pages = {642-651},
          doi = {10.1111/j.1365-2966.2011.18429.x},
archivePrefix = {arXiv},
       eprint = {1101.5092},
 primaryClass = {astro-ph.GA},
       adsurl = {https://ui.adsabs.harvard.edu/abs/2011MNRAS.414..642C}
}

@ARTICLE{magrini+09,
       author = {{Magrini}, Laura and {Stanghellini}, Letizia and {Villaver}, Eva},
        title = "{The Planetary Nebula Population of M33 and its Metallicity Gradient: A Look Into the Galaxy's Distant Past}",
      journal = {\apj},
         year = 2009,
        month = may,
       volume = {696},
       number = {1},
        pages = {729-740},
          doi = {10.1088/0004-637X/696/1/729},
archivePrefix = {arXiv},
       eprint = {0901.2273},
 primaryClass = {astro-ph.SR},
       adsurl = {https://ui.adsabs.harvard.edu/abs/2009ApJ...696..729M}
}

@ARTICLE{perinotto+04,
       author = {{Perinotto}, M. and {Morbidelli}, L. and {Scatarzi}, A.},
        title = "{A reanalysis of chemical abundances in galactic PNe and comparison with theoretical predictions}",
      journal = {\mnras},
         year = 2004,
        month = apr,
       volume = {349},
       number = {3},
        pages = {793-815},
          doi = {10.1111/j.1365-2966.2004.07470.x},
       adsurl = {https://ui.adsabs.harvard.edu/abs/2004MNRAS.349..793P}
}

@ARTICLE{barker07,
       author = {{Barker}, Michael K. and {Sarajedini}, Ata and {Geisler}, Doug and {Harding}, Paul and {Schommer}, Robert},
        title = "{The Stellar Populations in the Outer Regions of M33. III. Star Formation History}",
      journal = {\aj},
         year = 2007,
        month = mar,
       volume = {133},
       number = {3},
        pages = {1138-1160},
          doi = {10.1086/511186},
archivePrefix = {arXiv},
       eprint = {astro-ph/0611892},
 primaryClass = {astro-ph},
       adsurl = {https://ui.adsabs.harvard.edu/abs/2007AJ....133.1138B}
}

@ARTICLE{McConnachie06,
       author = {{McConnachie}, Alan W. and {Chapman}, Scott C. and {Ibata}, Rodrigo A. and {Ferguson}, Annette M.~N. and {Irwin}, Mike J. and {Lewis}, Geraint F. and {Tanvir}, Nial R. and {Martin}, Nicolas},
        title = "{The Stellar Halo and Outer Disk of M33}",
      journal = {\apjl},
         year = 2006,
        month = aug,
       volume = {647},
       number = {1},
        pages = {L25-L28},
          doi = {10.1086/507299},
archivePrefix = {arXiv},
       eprint = {astro-ph/0606728},
 primaryClass = {astro-ph},
       adsurl = {https://ui.adsabs.harvard.edu/abs/2006ApJ...647L..25M}
}

@MISC{williams21,
       author = {{Williams}, Benjamin F. and {Bell}, Eric F. and {Boyer}, Martha L. and {D'Souza}, Richard and {Dalcanton}, Julianne and {Diaz Rodriguez}, Mariangelly and {Dolphin}, Andrew Eugene and {Durbin}, Meredith and {Gilbert}, Karoline and {Girardi}, Leo and {Gordon}, Karl D. and {Guhathakurta}, Puragra and {Hammer}, Francois and {Johnson}, Lent Clifton and {Lang}, Dustin and {Lauer}, Tod R. and {Lazzarini}, Margaret and {Murphy}, Jeremiah and {Patel}, Ekta and {Quirk}, Amanda and {Roman-Duval}, Julia Christine and {Sanderson}, Robyn and {Seth}, Anil C. and {Smercina}, Adam and {Weisz}, Daniel R.},
        title = "{The Panchromatic Hubble Andromeda Southern Treasury (PHAST)}",
 howpublished = {HST Proposal. Cycle 29, ID. \#16778},
         year = 2021,
        month = jun,
        pages = {16778},
       adsurl = {https://ui.adsabs.harvard.edu/abs/2021hst..prop16778W}
}

@ARTICLE{gutierrez-soto25,
       author = {{Guti{\'e}rrez-Soto}, L.~A. and {Lopes de Oliveira}, R. and {Akras}, S. and {Gon{\c{c}}alves}, D.~R. and {Lomel{\'\i}-N{\'u}{\~n}ez}, L.~F. and {Mendes de Oliveira}, C. and {Telles}, E. and {Alvarez-Candal}, A. and {Borges Fernandes}, M. and {Daflon}, S. and {Ferreira Lopes}, C.~E. and {Grossi}, M. and {Hazarika}, D. and {Humire}, P.~K. and {Lima-Dias}, C. and {Lopes}, A.~R. and {Nilo Castell{\'o}n}, J.~L. and {Panda}, S. and {Kanaan}, A. and {Ribeiro}, T. and {Schoenell}, W.},
        title = "{Mapping H{\ensuremath{\alpha}} excess candidate point sources in the southern hemisphere using S-PLUS data}",
      journal = {\aap},
         year = 2025,
        month = mar,
       volume = {695},
          eid = {A104},
        pages = {A104},
          doi = {10.1051/0004-6361/202453167},
archivePrefix = {arXiv},
       eprint = {2501.16530},
 primaryClass = {astro-ph.GA},
       adsurl = {https://ui.adsabs.harvard.edu/abs/2025A&A...695A.104G}
}

@ARTICLE{lomeli25,
       author = {{Lomel{\'\i}-N{\'u}{\~n}ez}, Luis and {Cortesi}, A. and {Smith Castelli}, A.~V. and {Buzzo}, M.~L. and {Mayya}, Y.~D. and {Fragkou}, Vasiliki and {Alzate-Trujillo}, J.~A. and {Haack}, R.~F. and {Calder{\'o}n}, J.~P. and {Lopes}, A.~R. and {Hilker}, Michael and {Grossi}, M. and {Men{\'e}ndez-Delmestre}, Kar{\'\i}n and {Gon{\c{c}}alves}, Thiago S. and {Chies-Santos}, Ana L. and {Guti{\'e}rrez-Soto}, L.~A. and {Lima-Dias}, Ciria and {Werner}, S.~V. and {Humire}, Pedro K. and {Thom de Souza}, R.~C. and {Alvarez-Candal}, A. and {Panda}, Swayamtrupta and {Chaturvedi}, Avinash and {Telles}, E. and {Mendes de Oliveira}, C. and {Kanaan}, A. and {Ribeiro}, T. and {Schoenell}, W.},
        title = "{The S-PLUS Fornax Project (S+FP): Mapping Globular Clusters Systems within 5 Virial Radii around NGC 1399}",
      journal = {\aj},
         year = 2025,
        month = may,
       volume = {169},
       number = {5},
          eid = {263},
        pages = {263},
          doi = {10.3847/1538-3881/adbf0c},
archivePrefix = {arXiv},
       eprint = {2503.15657},
 primaryClass = {astro-ph.GA},
       adsurl = {https://ui.adsabs.harvard.edu/abs/2025AJ....169..263L}
}

@INPROCEEDINGS{oaj,

   author = {{Cenarro}, A.~J. and {Moles}, M. and {Mar{\'{\i}}n-Franch}, A. and

	{Crist{\'o}bal-Hornillos}, D. and {Yanes D{\'{\i}}az}, A. and

	{Ederoclite}, A. and {Varela}, J. and {V{\'a}zquez-Rami{\'o}}, H. and

	{Valdivielso}, L. and {Ben{\'{\i}}tez}, N. and {Cepa}, J. and

	{Dupke}, R. and {Fern{\'a}ndez-Soto}, A. and {Mendes de Oliveira}, C. and

	{Sodr{\'e}}, L. and {Taylor}, K. and {Rueda-Teruel}, S. and

	{Rueda-Teruel}, F. and {Luis-Simoes}, R. and {Chueca}, S. and

	{Ant{\'o}n}, J.~L. and {Bello}, R. and {D{\'{\i}}az-Mart{\'{\i}}n}, M.~C. and

	{Guill{\'e}n-Civera}, L. and {Hern{\'a}ndez-Fuertes}, J. and

	{Iglesias-Marzoa}, R. and {Jim{\'e}nez-Mej{\'{\i}}as}, D. and

	{Lasso-Cabrera}, N.~M. and {L{\'o}pez-Alegre}, G. and {L{\'o}pez-Sainz}, A. and

	{Rodr{\'{\i}}guez-Hern{\'a}ndez}, M.~A.~C. and {Su{\'a}rez}, O. and

	{Lamadrid}, J.~L. and {Ma{\'{\i}}cas}, N. and {Abril-Iba{\~n}ez}, J. and

	{Tilve}, V. and {Rodr{\'{\i}}guez-Llano}, S.},

	title = "{The Observatorio Astrof{\'{\i}}sico de Javalambre: current status, developments, operations, and strategies}",

booktitle = {Observatory Operations: Strategies, Processes, and Systems V},

 	year = 2014,

   series = {\procspie},

   volume = 9149,

	month = aug,

  	eid = {91491I},

	pages = {91491I},

  	doi = {10.1117/12.2055455},

   adsurl = {http://esoads.eso.org/abs/2014SPIE.9149E..1IC}

}

@INPROCEEDINGS{t80cam,

   	author = {{Mar{\'{\i}}n-Franch}, Antonio and {Taylor}, Keith and {Cenarro}, Javier and

     	{Cristobal-Hornillos}, David and {Moles}, Mariano},

    	title = "{T80Cam: a wide field camera for the J-PLUS survey}",

	booktitle = {IAU General Assembly},

     	year = "2015",

   	volume = {29},

    	month = "Aug",

      	eid = {2257381},

    	pages = {2257381},

   	adsurl = {https://ui.adsabs.harvard.edu/abs/2015IAUGA..2257381M}

}

@INPROCEEDINGS{upad,

   author = {{Crist{\'o}bal-Hornillos}, D. and {Gruel}, N. and {Varela}, J. and

	{L{\'o}pez-Sainz}, A. and {Ederoclite}, A. and {Moles}, M. and

	{Cenarro}, A.~J. and {Mar{\'{\i}}n-Franch}, A. and {Hern{\'a}ndez-Fuertes}, J. and

	{Yanes-D{\'{\i}}az}, A. and {Chueca}, S. and {Rueda-Teruel}, S. and

	{Rueda-Teruel}, F. and {Luis-Simoes}, R.},

	title = "{J-PAS data management pipeline and archiving}",

booktitle = {SPIE CS},

 	year = 2012,

   series = {},

   volume = 8451,

	month = sep,

  	eid = {845116},

  	doi = {10.1117/12.925431},

   adsurl = {http://esoads.eso.org/abs/2012SPIE.8451E..16C}

}

\end{document}